\documentclass[12pt]{iopart}

\usepackage{iopams}
\expandafter\let\csname equation*\endcsname\relax %I had to put in these two lines to use the ams math package without errors.
\expandafter\let\csname endequation*\endcsname\relax
\usepackage{amsmath}
\usepackage{graphicx}

\usepackage{hyperref}
\usepackage[separate-uncertainty=true,multi-part-units=single]{siunitx}
\usepackage[usenames,dvipsnames]{xcolor}

\usepackage{braket}

\newcommand{\lambdag}{\lambda_\mathrm{so}^\mathrm{g}}
\newcommand{\lambdau}{\lambda_\mathrm{so}^\mathrm{u}}
\newcommand{\lambdagu}{\lambda_\mathrm{so}^\mathrm{g,u}}
\newcommand{\lambdaug}{\lambda_\mathrm{so}^\mathrm{g,u}}

\newcommand{\Gammad}{\mathit{\gamma}_d}

\newcommand{\Gammaeup}{\mathit{\gamma}_+^\mathrm{u}}
\newcommand{\Gammaedown}{\mathit{\gamma}_-^\mathrm{u}}
\newcommand{\Gammagup}{\mathit{\gamma}_+^\mathrm{g}}

\newcommand{\gammar}{\mathit{\gamma}_\mathrm{r}}
\newcommand{\gammanr}{\mathit{\gamma}_\mathrm{nr}}

\newcommand{\doubEg}{{}^2\mathrm{E}_\mathrm{g}}
\newcommand{\doubEu}{{}^2\mathrm{E}_\mathrm{u}}

\newcommand{\Aone}{\mathrm{A}_1}

\newcommand{\A}{\mathrm{A}}
\newcommand{\E}{\mathrm{E}}

\newcommand{\NV}{\mathrm{NV}^-}
\newcommand{\nv}{\mathrm{NV}^-}
\newcommand{\siv}{\mathrm{SiV}^-}
\newcommand{\SiV}{\mathrm{SiV}^-}
\newcommand{\Dthreed}{\mathrm{D}_\mathrm{3d}}
\newcommand{\adagger}{a^\dagger}
\newcommand{\Tone}{\mathit{T}_1}

\begin{document}

\title[]{Electron-phonon processes of the silicon-vacancy centre in diamond}

\author{Kay D. Jahnke$^{1,}$\footnote{These authors contributed equally},
Alp Sipahigil$^{2,}$\footnotemark[\value{footnote}],
Jan M. Binder$^1$,
Marcus W. Doherty$^3$,
Mathias Metsch$^1$,
Lachlan J. Rogers$^1$\footnote{Corresponding author},
Neil B. Manson$^3$,
Mikhail D. Lukin$^2$ and
Fedor Jelezko$^1$
}

\address{${}^1$ Institute for Quantum Optics and IQST, Ulm University, Albert-Einstein-Allee 11, D-89081 Ulm, Germany}
\address{${}^2$ Department of Physics,	Harvard University,	17 Oxford Street, Cambridge, MA 02138, USA}
\address{${}^3$ Laser Physics Centre, Research School of Physics and Engineering, Australian National University, ACT 0200, Australia}
\ead{lachlan.j.rogers@quantum.diamonds}

\begin{abstract}
We investigate phonon induced electronic dynamics in the ground and excited states of the negatively charged silicon-vacancy ($\SiV$) centre in diamond. % and its optical transition. 
Optical transition line widths, transition wavelength and excited state lifetimes are measured for the temperature range \SIrange[range-phrase=--]{4}{350}{\kelvin}.
The ground state orbital relaxation rates are measured using time-resolved fluorescence techniques. 
A microscopic model of the thermal broadening in the excited and ground states of the $\SiV$ centre is developed. 
A vibronic process involving single-phonon transitions is found to determine orbital relaxation rates for both the ground and the excited states at cryogenic temperatures. 
We discuss the implications of our findings for coherence of qubit states in the ground states and propose methods to extend coherence times of $\SiV$ qubits.

\end{abstract}

\submitto{\NJP}

\maketitle

% just as a tool to help us understand the structure right now.
\tableofcontents

\section{Introduction}

Colour centres in diamond have emerged as attractive systems for applications in quantum metrology, quantum communication, and quantum information processing \cite{childress2014atomlike,pfaff2014unconditional,ladd2010quantum}.
Diamond has a large band gap which allows for optical control, and it can be synthesised with high purity enabling long coherence times as was demonstrated for nitrogen-vacancy ($\NV$) spin qubits \cite{balasubramanian2009ultralong}.
Among many colour centres in diamond \cite{aharonovich2011diamond,aharonovich2014diamond}, the negatively charged silicon-vacancy ($\siv$) centre stands out due to its desirable optical properties.
In particular, near transform limited photons can be created with high efficiency due to the strong zero-phonon line (ZPL) emission that constitutes $\sim$\SI{70}{\percent} of the total emission.
$\siv$ centres can also be created with a narrow inhomogeneous distribution that is comparable to the transform limited optical line width \cite{rogers2014multiple}.
These optical properties, due to the inversion symmetry of the system which suppresses effects of spectral diffusion, recently enabled demonstration of two-photon interference from separated emitters \cite{sipahigil2014indistinguishable} that is a key requirement for many quantum information processing protocols \cite{childress2005fault,lim2006repeat,obrien2007optical,knill2001scheme}.

Interfacing coherent optical transitions with long-lived spin qubits is a key challenge for quantum optics with solid state emitters \cite{togan2010quantum,gao2012observation,de2012quantum,chu2014coherent,faraon2012coupling}. This challenge may be addressed using optically accessible electronic spins in $\SiV$ centres \cite{muller2014optical}.
It has recently been demonstrated that coherent spin states can be prepared and read out optically \cite{rogers2014all-optical, pingault2014all}, although the spin coherence time was found to be limited by phonon-induced relaxation in the ground states \cite{rogers2014all-optical}.
Here we present the first systematic study of the electron-phonon interactions that are responsible for relaxation within the ground and excited states of the $\siv$ centre.
This is achieved by measuring the temperature dependence of numerous processes within the centre.
A comprehensive microscopic model is then developed to account for the observations.
In Section \ref{subsec:spin_discussion} we discuss the implications of these phonon processes for spin coherences in the $\SiV$ ground state, and identify approaches that could extend the spin coherences.

\begin{figure}[thb]
\flushright
\includegraphics{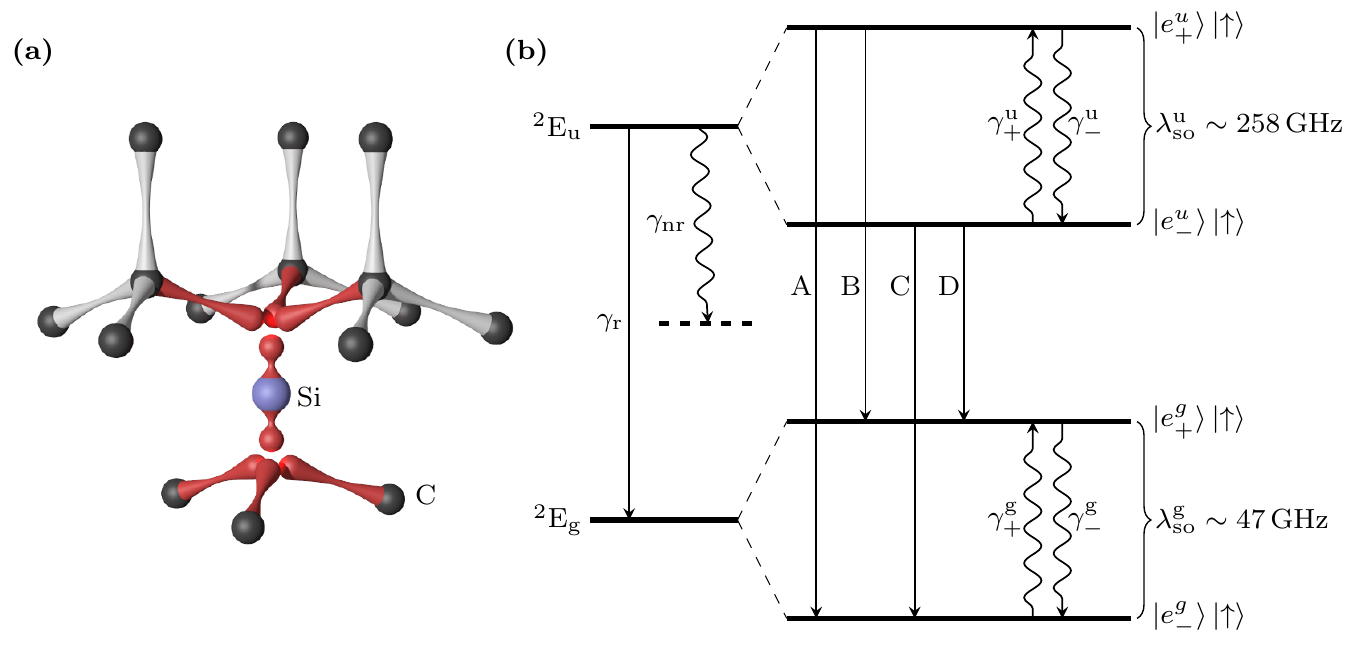}
\caption{
	Molecular structure and electronic dynamics of $\SiV$.
	(a) The $\SiV$ centre consists of a silicon atom centred between two neighbouring vacant lattice sites.	
	(b) The optical transitions are between states of E symmetry with opposite parity ($\doubEg ,\doubEu$).
	$\gammar$ ($\gammanr$) are radiative (non-radiative) decay rates out of the excited states.
	Straight (curved) lines denote the radiative (non-radiative) transitions.
	In both the ground and excited states, the four-fold degeneracy is partially lifted by the spin-orbit interaction $\lambdaug$ \cite{rogers2014electronic,hepp2014electronic}.
	Every level illustrated here is a spin-1/2 doublet (e.g. $\{\ket{e_-}\ket{\uparrow} ,\ket{e_+}\ket{\downarrow}\}$ for the lowest energy level), and for clarity only the spin-up levels are drawn.
	Implications of this study for the spin sublevels are discussed in Section \ref{subsec:spin_discussion}.
    The horizontal dashed line denotes the unidentified level (either an additional electronic level or excited vibrational state of $\doubEg $) involved in the non-radiative decay between the ground and excited states.
}
\label{fig:level}
\end{figure}

The $\SiV$ centre consists of an interstitial silicon atom in a split-vacancy configuration with $\Dthreed$ symmetry as illustrated in \autoref{fig:level}(a) \cite{goss1996twelveline}.
This symmetry gives rise to an electronic level structure consisting of ground ($\doubEg$) and excited ($\doubEu$) states that both have $\E$ symmetry and double orbital degeneracy.
The degenerate orbital states are occupied by a single hole with $S=1/2$ \cite{johansson2011optical,rogers2014electronic, hepp2014electronic,gali2013emphab}, leading to both orbital and spin degrees of freedom.
In the absence of off-axis strain or magnetic fields, the spin-orbit interaction ($\sim \lambdagu S_zL_z$) determines the eigenstates with well defined orbital and spin angular momentum \cite{hepp2014electronic}.
Optical and phononic transitions between these eigenstates couple only to the orbital degree of freedom and are spin conserving.
We therefore focus on the orbital dynamics within the ground and excited states, which can each be described as an effective two-level system consisting of two orbital states $\{ |L_z=\pm1 \rangle =|e^{g,u}_\pm\rangle\}$ for a given spin projection as shown in \autoref{fig:level}(b)\cite{hepp2014electronic}.
Phonons can introduce vibronic coupling between $|e_+\rangle$ and $|e_-\rangle$ orbitals, resulting in population transfer between orbitals at rates
$\gamma^{g,u}_{+,-}$ \cite{ham1965dynamical,fischer1984vibronic}.
This effect, also called the dynamic Jahn-Teller effect, has been observed in the excited states of the $\nv$ centre \cite{fu2009observation,goldman2014phonon} where a similar orbital degeneracy is present.

\section{Experimental results}
\label{sec:experimental}
\subsection{Excited states}
\label{sec:excited}

The spin-orbit interaction results in four optical dipole transitions, labelled A--D in order of increasing wavelength, centred around $\sim$\SI{737}{\nano\meter} at cryogenic temperatures \cite{sternschulte1994luminescence,clark1995silicon,neu2011single}.
At liquid helium temperatures, the optical line widths are broader for transitions A and B than for the lower energy transitions C and D \cite{rogers2014multiple}.
This was attributed to thermal relaxation reducing the effective lifetime of the upper branch via the decay rate $\Gammaedown$, which is faster than $\Gammaeup$ by the Boltzmann factor $\Gammaedown = \Gammaeup \exp( \lambdau / k_\mathrm{B} T )$ \cite{sternschulte1994luminescence, clark1995silicon} .
%AS: I think Sternschulte is appropriate. ensembles are enough to conclude this.

To probe the microscopic mechanism of the orbital relaxation in the excited states, the temperature dependence of the line width of transition D was measured for individual $\SiV$ centres.
Since these $\SiV$ centres exhibit negligible spectral diffusion \cite{rogers2014multiple}, the measured optical line widths $\Gamma (T)$ correspond to homogeneous broadening mechanisms associated with depolarisation and dephasing: $\Gamma (T) = \gammar + \gammanr(T) + \Gammaeup(T) + \Gammad(T)$.
The non-radiative decay rate $\gammanr(T)$ has a very weak temperature dependence, as discussed later in Section \ref{sec:optical_transition}, leading to a small contribution compared to the other rates for all temperature regions of interest.
The most significant temperature dependence comes from the relaxation rates within the excited states: $\Gammaeup(T)$ and $\Gammad(T)$.
As will be shown in Section \ref{sec:first-order_transitions}, the optical transition line width is dominated by relaxation rates in the excited states with little contribution from the ground states.

\begin{figure}[thb]
\flushright
\includegraphics{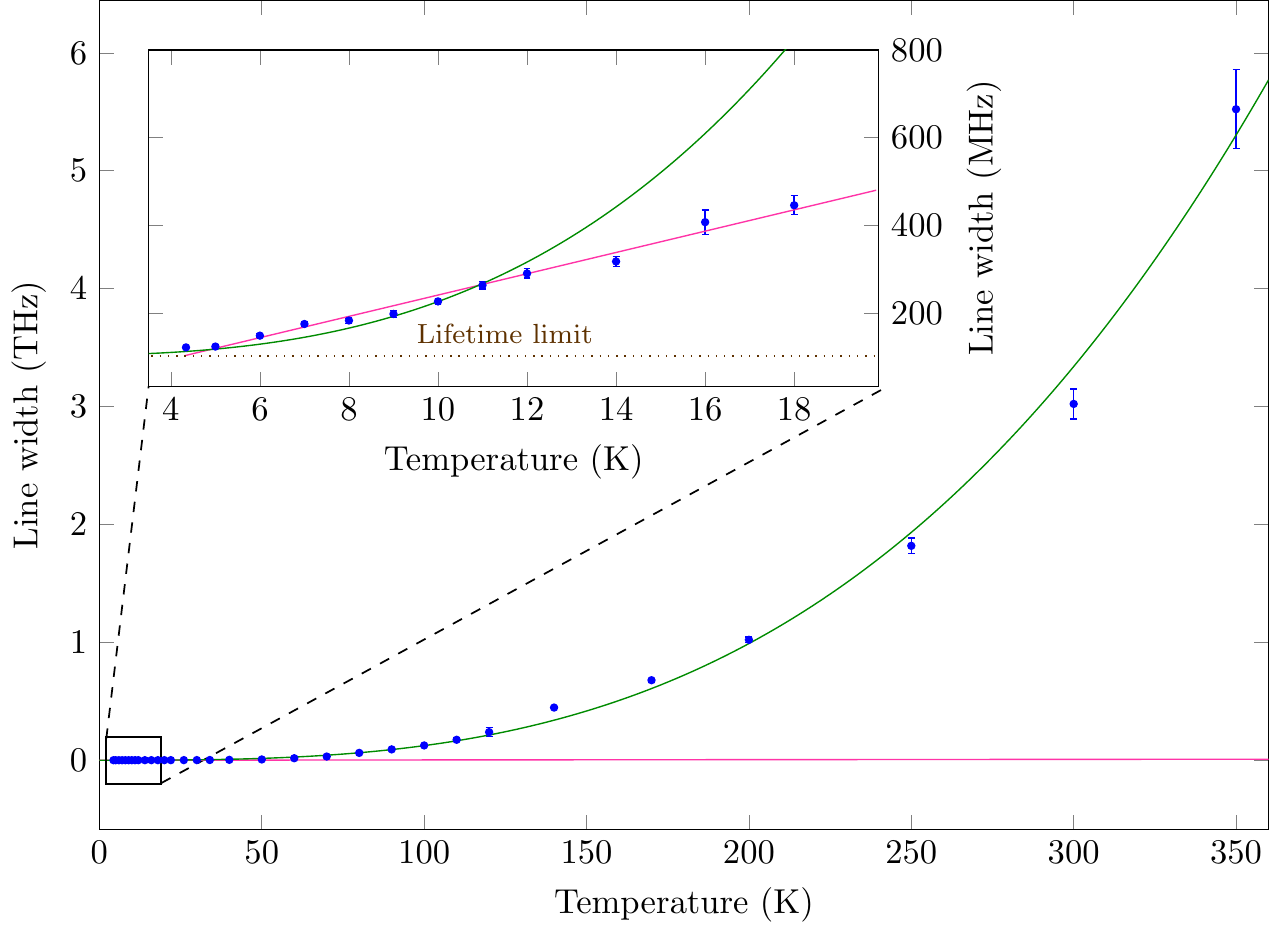}
\caption{
	Line width of transition D measured for different temperatures.
	Each line width was determined by Lorentzian fits for multiple sites.
	The green fit corresponds to a cubic scaling over the high temperature range ($>\SI{70}{\kelvin}$) after the spectrum has merged to two peaks and one peak at $>\SI{120}{\kelvin}$.
	At low temperatures ($<20$\,K) the pink fit represents a linear scaling seen in the inset.
	For the temperature range in between these two regimes the scaling of the line width crosses over from linear to cubic behaviour.
}
\label{fig:linewidth}
\end{figure}

For temperatures between \SIrange{4}{50}{\kelvin}, the optical transition line width was measured using photoluminescence excitation in a continuous flow cryostat, where a weak probe laser was scanned across transition D and fluorescence in the phonon-sideband (PSB) was detected.
At higher temperatures,  \SI{532}{\nano\meter} excitation was used and emission line widths were measured with a spectrometer (Princeton Instruments Acton 2500 equipped with a Pixis 100 cooled CCD-array and a \SI{1596}{\per\milli\meter} grating) giving a resolution of \SI{16}{\giga\hertz}.
To measure fundamental properties of the $\siv$ centre, a bulk diamond sample containing highly uniform defect sites and low strain was used in these experiments.
It was a low strain HPHT type-IIa diamond with a \{100\} surface on which a \SI{60}{\micro\meter} layer incorporating $\siv$ was created by microwave-plasma chemical-vapour-deposition (MPCVD).
This sample was used in previous publications \cite{rogers2014multiple,sipahigil2014indistinguishable} and shows a narrow inhomogeneous distribution for the $\siv$ optical transitions.

\autoref{fig:linewidth} shows the full width at half maximum (FWHM) line widths (determined from Lorentzian fits) measured for single $\siv$ sites in a $200\times 200$\,\SI{}{\micro\meter}${}^2$ region containing ${}^{28}$Si. 
Above $\sim\SI{70}{K}$ the line width scales as the cube of the temperature ($\Gamma = (103 + 0.12  \cdot (T/\mathrm{K})^3)$ MHz).
However for low temperatures ($ < \SI{20}{K}$), the behaviour deviates from $T^3$ and is better approximated by a linear dependence on temperature ($\Gamma = (-1.05 + 24.26 \cdot (T/\mathrm{K}))$  MHz) saturating at about \SI{4}{\kelvin} to the lifetime limited line width.
Early studies on nanodiamonds have measured the $T^3$ dependence of the line width on temperature, but were not able to resolve this linear contribution due to a combination of inhomogeneous broadening and spectral resolution limits \cite{neu2013low}.
It is shown in Section \ref{sec:model} that the observed $T$ and $T^3$ mechanisms result from first- and second-order transitions due to linear electron-phonon interactions with $\E$-symmetric phonon modes.

\subsection{Ground states}
\label{sec:exp_ground_states}

\begin{figure}
\flushright
\includegraphics{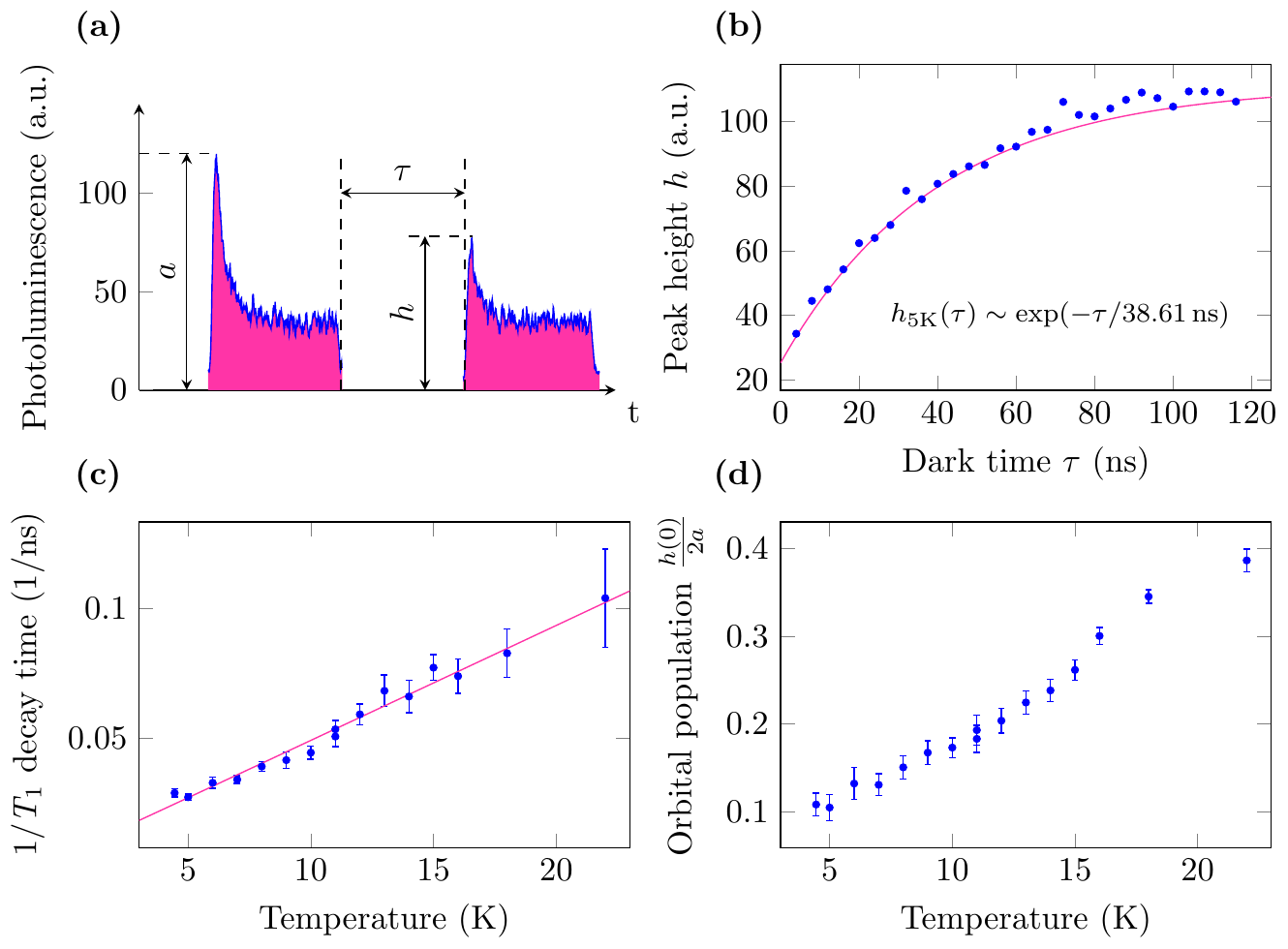}
\caption{
	Ground state orbital relaxation time ($\Tone$) measurements.
	(a) The pulse sequence used to measure the $\Tone$ of the ground states.
	A single laser was amplitude modulated to pump (first pulse) and probe (second pulse) transition D at each temperature.
	Photoluminescence (PL) intensity corresponds to the bright state population.
	(b) The height $h$ of the leading edge peak  plotted for different wait times $\tau$ between the pulses.
	An exponential fit to the recovery of the height gives the orbital ground state relaxation time $\Tone$.
	(c) Measured orbital relaxation rate, $\Gammagup$, as a function of temperature.
	A fit (pink) to a single-phonon relaxation model (Section \ref{sec:first-order_transitions}) shows good agreement with the data by introducing an offset on the temperature ($1/T_1=0.0099 \cdot n(\Delta,T-\SI{2.26}{\kelvin})$).
	(d) Bright state population, $h(0)/2a$, after optical pumping shown for different temperatures.
}
\label{fig:ground_T1}
\end{figure}

Relaxation within the ground state doublet, $\Gammagup$ in \autoref{fig:level} (b), was probed directly using pulsed excitation and time-resolved fluorescence measurements.
Transitions C and D form an optical $\Lambda$-system which allows ground state populations to be optically pumped.
For these experiments a second diamond sample was used in which the properties and orientation of the $\siv$ centres were known from earlier studies \cite{rogers2014electronic}.
The sample is a low strain high-pressure high-temperature (HPHT) diamond observed through a \{111\} surface with a low density of in-grown $\siv$ centres.
A laser was tuned to transition D and \SI{80}{\nano\second} pulses were generated using an electro-optical amplitude modulator with a measured extinction ratio of up to \SI{20}{\deci\bel}.
The signal was detected by counting the photon arrival times in relation to the laser pulses using a time-tagged data acquisition card (FAST ComTec MCS26A) giving a time resolution of up to \SI{200}{\pico\second}.

At the start of each laser pulse we observed a fluorescence peak that decayed to a steady state level.
For the first laser pulse, the peak height $a$ corresponds to the thermal population in the bright state ($|e^g_+\rangle\ket{\uparrow}$ or $|e^g_-\rangle\ket{\downarrow}$ ) which is $\sim \SI{50}{\percent}$ for the temperatures in our measurements.
The decay of this initial peak when the laser is on corresponds to optical pumping into the dark state ($|e^g_-\rangle\ket{\uparrow}$ or $|e^g_+\rangle\ket{\downarrow}$).
After a dark interval $\tau$, the dark state relaxes back to the bright state, leading to a recovery of peak height $h$ for subsequent pulses.
The peak height, $h(\tau)$, exhibits a simple exponential recovery indicating a single characteristic relaxation time, $\Tone$, as shown in \autoref{fig:ground_T1}(a,b).
This measurement was repeated for a single $\SiV$ centre at various temperatures between \SI{4.5}{\kelvin} and \SI{22}{\kelvin} and the relaxation rate was found to scale linearly with temperature (\autoref{fig:ground_T1})(c).
The longest $\Tone$ time was measured at the lowest temperatures to be $\Tone(\SI{5}{\kelvin})= \SI{39\pm1}{\nano\second}$.

We note that the steady state fluorescence level under laser excitation, $h(0)$, is determined by a competition between the optical pumping rate and thermalization rates ($1/\Tone$).
With increasing temperature, the thermalization rate increases (\autoref{fig:ground_T1}(c)) while the optical pumping rate remains nearly constant at saturation.
This leads to a reduced measured peak contrast ($h(0)/2a$, \autoref{fig:ground_T1}(d)) with increasing temperature showing that the ground state of the $\siv$ centre cannot be polarised at elevated temperatures.

\subsection{Excited state lifetimes}
\label{sec:optical_transition}

The results presented so far have only highlighted the processes within the ground and excited state doublets.
The transition rates from excited to ground states also have a temperature dependence which can be probed by measuring the fluorescence lifetime of the excited state as a function of temperature.
Previous experiments have reported excited state lifetimes in the $\sim\,$\SIrange{1}{4}{\nano\second} range along with various estimates of the quantum yield  \cite{sternschulte1994luminescence, neu2011single, neu2013low, lee2012coupling,  moller2014deterministic, rogers2014multiple}.
\autoref{fig:level}(b) shows potential radiative ($\gammar$) and non-radiative ($\gammanr$) processes taking place at the optical energy scale that determine the excited state lifetimes and the quantum yield.
The total decay rate from the excited states, $\gamma_t (T)=1/\tau_0(T)=\gammar +\gammanr (T)$, is a combination of a constant radiative ($\gammar$) and a temperature dependent non-radiative rate ($\gammanr (T)$).

\begin{figure}
\flushright
\includegraphics{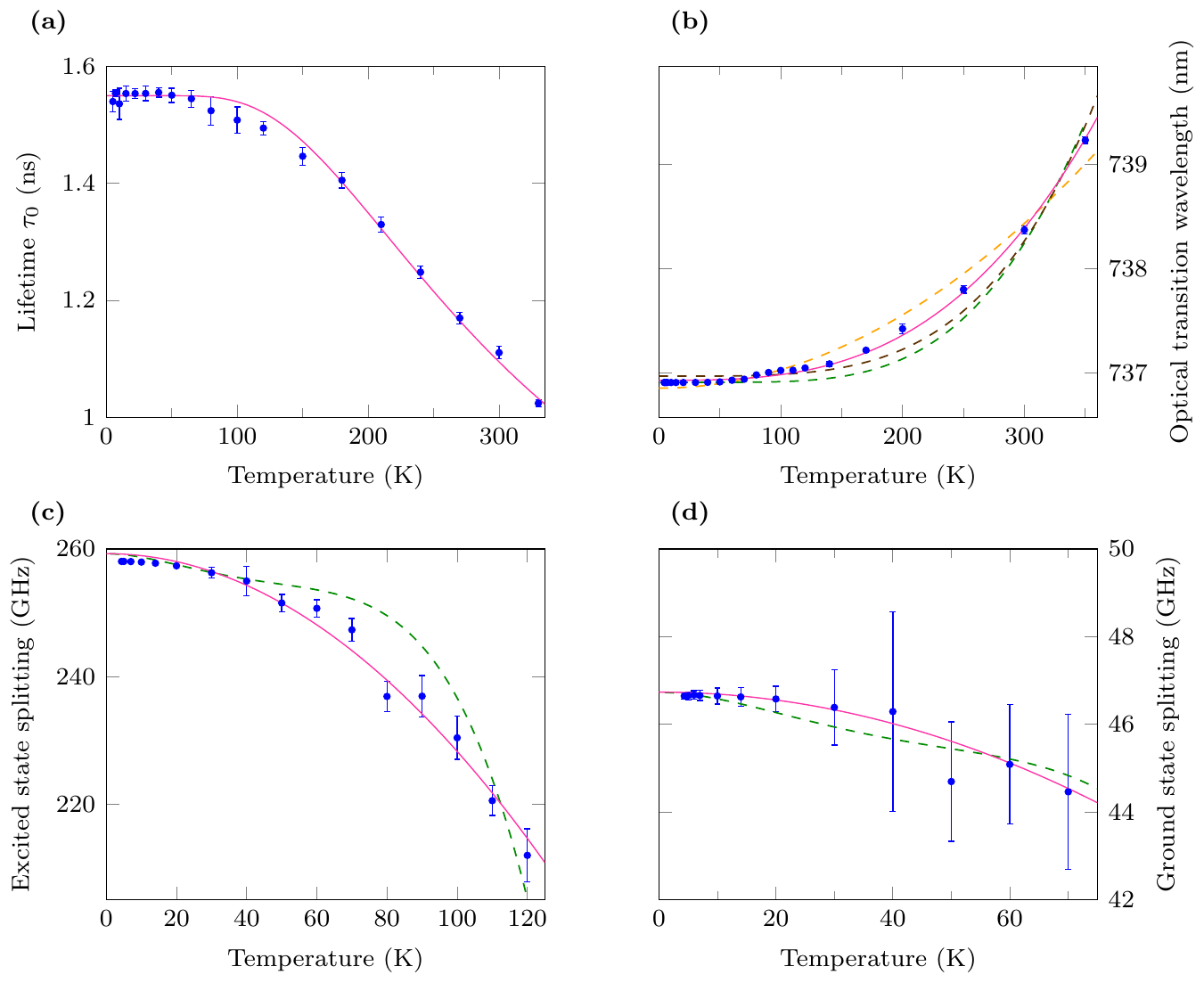}
\caption{
	(a) Fluorescence lifetime ($\tau_0$) of the excited states as a function of temperature.
	At each temperature, $\tau_0(T)$ was measured for 10 emitters.
	The error bars denote the standard deviation of the $\tau_0 (T)$ distribution.
	The fit (magenta) line corresponds to the Mott-Seitz model.
	(b) Optical transition wavelength of transition C determined from Lorentzian fits to the spectrum and excitation scans.
	A cubic (magenta) dependence on temperature ($\sim T^3$) is in good agreement with the data, unlike quadratic (dashed, orange) and quartic (dashed, brown) fits.
	(c,d) The measured excited and ground state splittings as a function of temperature. The quadratic fits  based on the model in Section \ref{sec:wavelength} are shown in magenta.
	The dashed green lines in (b), (c) and (d) are the best fits obtained using the pure thermal expansion mechanism explained in the text.
}
\label{fig:optical_transition}
\end{figure}

In an attempt to identify the non-radiative process, we measured the lifetime of the $\doubEu$ excited states as a function of temperature from \SIrange{5}{350}{\kelvin}.
At each temperature, 10 separate single $\siv$ centres were excited using a pulsed \SI{532}{\nano\meter} laser with a \SI{80}{\mega\hertz} repetition rate and the measured time traces were fitted using a single exponential decay.
The measured temperature dependence of $\tau_0(T)=1/\gamma_t (T)$ is shown in \autoref{fig:optical_transition}(a).
The excited state lifetime was found to increase as temperature was decreased down to \SI{50}{\kelvin}, where it saturated to a constant level.
These results suggest there is a finite non-radiative rate $\gammanr$ at room temperature, while the saturation below \SI{50}{\kelvin} does not necessarily imply $\gammanr (T<\SI{50}{\kelvin})=0$ as there might still be a finite spontaneous non-radiative rate at zero temperature.
The observed temperature dependence in \autoref{fig:optical_transition}(a) can be described by the Mott-Seitz model for non-radiative relaxation, $\tau_0(T)=\tau_0(T=\SI{0}{\kelvin})(1+\alpha e^{-\frac{\Delta E}{k_BT}})^{-1}$, with an activation energy of $\Delta E=\SI{55\pm2}{\milli\electronvolt}$ and $\alpha=3.3\pm0.3$ \cite{toyli2012measurement}.
Our measurements do not, however, provide enough information to distinguish whether the system decays from $\doubEu$ directly to a higher vibrational state of $\doubEg$, or to an unidentified electronic level closer to $\doubEu$ in energy.
Whilst there exists some \textit{ab initio} \cite{gali2013emphab} and experimental \cite{neu2012electronic} evidence of an additional electronic level below the excited $\doubEu$ level, this evidence conflicts with the simple molecular orbital model of the centre’s electronic structure \cite{goss1996twelveline,johansson2011optical,rogers2014electronic,hepp2014electronic}, which predicts no such additional level.
Future studies involving spectroscopy of the $\doubEg\rightarrow\doubEu$ absorption PSB and single-shot readout capability of $\siv$ electronic states might help identify the relaxation paths from the $\doubEu$ and dark states of $\siv$ centres \cite{waldherr2011dark}.

\subsection{Optical line positions}
\label{sec:wavelength}
The line positions of all four optical transitions were determined using the Lorentzian fits to the measurements described in Section \ref{sec:excited}.
The spectrometer was calibrated with respect to a wavemeter, allowing us to consistently reproduce transition wavelengths across the entire temperature range.
For simplicity, only transition C is shown in \autoref{fig:optical_transition}(b).
Fitting with a free temperature exponent results in $\Delta\lambda=19.2\cdot (T/\mathrm{K})^{2.78 \pm 0.05}$, in close agreement with a cubic temperature dependence.
\autoref{fig:optical_transition}(b) compares fits of the form $\Delta\lambda  \sim T^\alpha$ for $\alpha=2,3,4$ as well as a model based on thermal expansion described below.
Our observation of $T^3$ scaling differs marginally from earlier measurements made on nanodiamonds \cite{neu2013low}.
For temperatures at which the linewidth was narrow enough to resolve individual transitions, the ground and excited state splittings could also be obtained from the spectrum.
We observe that the measured splittings, which correspond to the spin-orbit interaction at low temperature, are reduced with increasing temperature for both the excited (\autoref{fig:optical_transition}(c)) and ground (\autoref{fig:optical_transition}(c)) states.

In diamond, the temperature shifts of optical lines have two distinct origins: thermal expansion and electron-phonon interactions \cite{davies1974vibronic, doherty2014temperature}.
The shift of the transition energy due to thermal expansion has the form $\delta E_\mathrm{exp.}(T)=A \cdot P(T)$, where $A$ is the hydrostatic pressure shift of the transition energy, $P(T)=-B\int_0^T e(x)dx$ is the negative pressure of thermal expansion, $B$ is the diamond bulk modulus and $e(T)$ is the bulk thermal expansion coefficient  \cite{davies1974vibronic, doherty2014temperature}.
Whilst $B$ and $e(T)$ are well-known for diamond  \cite{sato2002thermal}, the pressure shift $A$ of the $\siv$ optical transition has not been measured.
The dashed green line in \autoref{fig:optical_transition}(b) is the best fit of the line shift obtained using the single fit parameter $A$ of the thermal expansion mechanism, and it is clear this does not account for the observed shift.
The shift of the transition energy due to electron-phonon interactions typically arises from quadratic interactions with $\Aone$-symmetric phonon modes and produces a $T^4$ dependence \cite{maradudin1966solid}, which is also inconsistent with our observations.
Furthermore, a linear combination of shifts caused by these two mechanisms is not able to produce a good fit to our observations.
It is shown in Section \ref{sec:model} that the atypical $T^3$ shift arises from second-order linear interactions with $\E$-symmetric phonon modes.

The temperature reductions of the ground and excited state splittings can also arise from thermal expansion and electron-phonon interactions \cite{doherty2014temperature}.
The dashed green lines in \autoref{fig:optical_transition}(c) and (d) are the best thermal expansion fits obtained by introducing pressure shift parameters of the spin-orbit splittings, and as above it is clear, at least for \autoref{fig:optical_transition}(c), that another mechanism must be involved in the reduction of the splittings.
Similarly, the $T^4$ dependence of the quadratic interactions with $\Aone$-symmetric phonon modes and its linear combination with the thermal expansion shift do not satisfactorily fit the observations.
We will show in the next section that the $T^2$ dependence of the spin-orbit splittings are also consequences of second-order linear interactions with $\E$-symmetric phonon modes.

\section{Microscopic model of the electron-phonon processes}
\label{sec:model}
\begin{figure}
\flushright
\includegraphics[width=134.6mm]{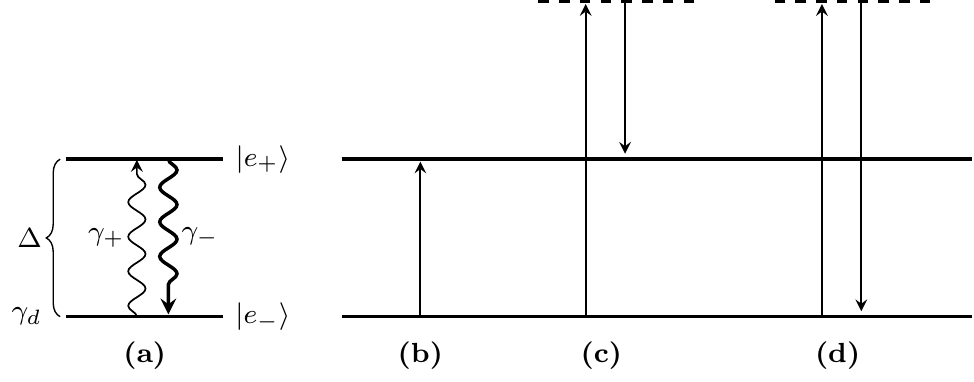}
\caption{
	Electron-phonon processes within the ground and excited states.
	(a) The relevant rates in the problem are $\gamma_+$ and $\gamma_-$, which denote the transition rates between states that determine the orbital $\Tone$ relaxation time, and $\gamma_d$ which denotes the rate of a pure dephasing process.
	These rates can originate from a (b) single-phonon direct process; (c) two-phonon Raman process; or (d) two-phonon elastic scattering process.
	These processes are analogous to (b) resonant absorption, (c) Raman transitions, and (d) AC Stark shift in atomic physics.
}
\label{fig:phonon_process}
\end{figure}

In this section we develop a microscopic model of the electron-phonon processes within the ground and excited electronic levels that are summarised in \autoref{fig:phonon_process}. 
The model successfully describes the observed temperature variations of optical line width, line position and spin-orbit splittings. 
In each case, the electron-phonon processes are consequences of the linear Jahn-Teller interaction between the $\E$-symmetric electronic states and $\E$-symmetric acoustic phonon modes \cite{fischer1984vibronic,maze2011properties,doherty2011negatively}.

As discussed in the introduction, spin projection and orbital angular momentum are good quantum numbers and can be treated separately in the situation of low magnetic fields and strain \cite{hepp2014electronic}.
Since the optical transitions and electron-phonon interactions are spin conserving, we can focus on the orbital degrees of freedom of the ground and excited levels.
For a given spin state, the effective zero-field orbital Hamiltonian takes the following form for both the ground and excited levels
\begin{equation}
H_0=\pm\frac{1}{2}\hbar \Delta \sigma_z\,,
\label{eq:zeroFieldHamiltonian}
\end{equation}
where $\sigma_z$ is the usual Pauli operator for orbital states in the $\{\ket{e_+}, \ket{e_-} \}$ basis, $\hbar\Delta$ is the magnitude of the relevant spin-orbit splitting, which is $+\hbar\Delta$ for $\ket{\uparrow}$ and $-\hbar\Delta$ for $\ket{\downarrow}$.

\subsection{Electron-phonon interaction}
\label{sec:electron-phonon_interaction}

For the $\siv$ centre the interaction between the $\E$-symmetric orbital states $\{\ket{e_+}, \ket{e_-} \}$ and phonon modes of $\E$ symmetry is described most easily if the $\E$ modes are linearly transformed to be circularly polarised.
With this transformation, the phonon Hamiltonian and the linear electron-phonon interaction are

\begin{eqnarray}
\mathrm{\hat{H}_E}&=\sum_{p,k}{\hbar\,\omega_k \,\adagger_{p,k}a_{p,k}}\\
\mathrm{\hat{V}_E}&=\sum_{k}\hbar \, \chi_k [\sigma_+ (a_{-,k}+\adagger_{-,k}) + \sigma_- (a_{+,k}+\adagger_{+,k}) ]\,,
\label{eq:electronPhonon}
\end{eqnarray}
where $\chi_k$ is the interaction frequency for a single phonon, $\sigma_+$ ($\sigma_-$) is the raising (lowering) operator for orbital states, and $a_{p,k}^\dagger$ ($a_{p,k}$) is the creation (annihilation) operator for phonons with polarisation $p=\{-,+\}$ and wavevector $k$.
For acoustic modes in bulk diamond, the interaction frequency and density of modes are approximately $\overline{|\chi_k(\omega)|^2}\approx \chi\,\omega$ and $\rho(\omega)=\rho\, \omega^2$, respectively, where the overbar denotes the average over all modes with frequency $\omega_k=\omega$ and $\chi$ and $\rho$ are proportionality constants \cite{fu2009observation,abtew2011dynamic}.
Note that interactions with $\A$-symmetric modes have  not been included as they do not couple the states within the ground and excited electronic levels.

\subsection{First-order electron-phonon transitions}
\label{sec:first-order_transitions}

Treating $\hat{V}_\mathrm{E}$ as a time-dependent perturbation, the first-order transitions between the orbital states involve the absorption or emission of a single $\E$ phonon whose frequency is resonant with the splitting $\Delta$ (see \autoref{fig:phonon_process} (b)).
The corresponding transition rates are
\begin{eqnarray}
\gamma_+&= 2\pi \sum_{k}{ n_{-,k} |\chi_k|^2 \delta(\Delta-\omega_{k})} \nonumber\\
\gamma_-&= 2\pi \sum_{k}{ (n_{+,k} +1)|\chi_k|^2 \delta(\Delta-\omega_{k})}\,,
\label{eq:firstorder}
\end{eqnarray}

\noindent where $n_{p,k}$ is the occupation of the phonon mode with polarisation p and wavector k.
Assuming acoustic phonons, performing the thermal average over initial states and the sum over all final states leads to
\begin{eqnarray}
\gamma_+&= 2\pi \chi \, \rho\, \Delta^3 n(\Delta,T) \nonumber\\
\gamma_-&= 2\pi \chi \, \rho\, \Delta^3 [n(\Delta,T)+1]\,.
\label{eq:firstOrderRate}
\end{eqnarray}
For temperatures $T>\hbar \Delta/k_\mathrm{B}$, Eq. (\ref{eq:firstOrderRate}) can be approximated by a single relaxation rate with a linear temperature dependence
\begin{equation}
\gamma_+ \approx \gamma_- \approx \frac{2\pi}{\hbar}\chi \rho \Delta^2k_\mathrm{B}T\,.
\label{eq:firstOrderApprox}
\end{equation}

Hence, the one-phonon transitions lead to the relaxation of population between the orbital states as well as the dephasing of the states that are linearly dependent on temperature.
%
%\todo{Although acoustic phonons have odd parity, this does not prevent a single-phonon transition between the two orbital branches of matching parity.  By considering phonons with well defined E symmetry, the modes are implicitly transformed into linear combinations of acoustic modes with the same frequency, but different polarizations and k vector directions.}
%
The measurements presented in \autoref{fig:linewidth} and \autoref{fig:ground_T1} demonstrated a clear linear dependence of broadening for temperatures below \SI{20}{\kelvin}, but greater than the spin orbit splitting ($T>\hbar \Delta/k_\mathrm{B}\sim\SI{2.4}{\kelvin}$).
We therefore conclude that the relaxation mechanisms are dominated by a resonant single phonon process at liquid helium temperatures for both the ground and the excited states.
Eq. (\ref{eq:firstOrderApprox}) also shows that the relaxation rate is $\sim\Delta^2$, where $\Delta$ is the spin-orbit splitting in the zero-field limit.
The $\Delta^2$ scaling explains why the phonon relaxation processes are much faster in the excited levels for which the splittings are larger compared with the ground states.

\subsection{Second-order electron-phonon transitions}
\label{sec:second-order_transitions}
It was seen in \autoref{fig:linewidth} that the line broadening deviated from its linear temperature dependence above about \SI{20}{\kelvin} ($T\gg\hbar \Delta/k_\mathrm{B}$), suggesting that higher order processes involving two phonons start dominating the relaxation rates.
Given the form of the electron-phonon interaction in Eq. (\ref{eq:electronPhonon}), the only allowed two-phonon processes are those where the initial and final orbital states are identical.
Therefore, the inelastic Raman-type scattering processes [\autoref{fig:phonon_process}(c)] that are dominant for $\nv$ centres \cite{fu2009observation}, are suppressed in $\siv$ and the elastic Raman-type scattering processes [\autoref{fig:phonon_process}(d)] dominate instead.
The elastic scattering rate for $|e_-\rangle$ is
\begin{eqnarray}
\gamma_{d-}= 2\pi \hbar^2 \sum_{k,q}  &n_{-,k } \,(n_{+,q} +1 ) |\chi_{k}|^2 |\chi_{q}|^2   \nonumber\\
&\left| \frac{1}{\Delta-\omega_{k}} +\frac{1}{\Delta+\omega_{k}} \right| ^2 \delta(\Delta-\omega_{k}+\omega_{q})\,.
\end{eqnarray}
\noindent Performing the thermal average over the initial states and the sum over all final states leads to
\begin{eqnarray}
\gamma_{d-}= 2\pi \hbar^2 \int_0^\Omega &n(\Delta+\omega,T) (n(\omega,T)+1)
\overline{\left| \chi_{k} (\Delta+\omega) \right|^2}\,\overline{\left| \chi_{q} (\omega) \right|^2} \nonumber\\
&\left| \frac{1}{-\omega}+\frac{1}{\Delta+\omega}\right|^2 \rho(\Delta+\omega)\rho(\omega)\textrm{d}\omega\,,
\end{eqnarray}

\noindent where $\Omega$ is the Debye frequency of diamond.
Assuming acoustic modes and that the temperatures are such that only modes with frequencies $\Omega\gg\omega\gg\Delta$ contribute significantly to the integral, to lowest order in $\Delta$, the rates become
\begin{eqnarray}
\gamma_{d-}\approx\gamma_{d+}\approx &2\pi \hbar^2\Delta^2\chi^2\rho^2\int_0^\infty n(\omega,T)(n(\omega,T)+1)\omega^2\textrm{d}\omega \nonumber\\
&=\frac{2\pi^3}{3\hbar}\Delta^2\chi^2\rho^2k_\mathrm{B}^3T^3\,.
\end{eqnarray}

Hence, the two-phonon elastic scattering process contribute to the dephasing of the orbital states and have rates that are proportional to $\sim T^3$, matching the observed line width behaviour in \autoref{fig:linewidth}.
Therefore our microscopic model shows perfect agreement with the measurements and we can understand the orbital relaxation process as a combination of a single phonon mixing between the orbital states and a two-phonon dephasing process.

For $\siv$ centres under high strain (larger than the spin-orbit interaction), the orbital eigenstates $\{|e_x\rangle, |e_y\rangle \}$ no longer have well defined angular momentum.
Under such conditions, the inelastic Raman process shown in \autoref{fig:phonon_process}(c)  becomes allowed, which results in a competing orbital relaxation rate that scales as $\sim T^5$.

\subsection{Spin-orbit splitting shifts}
\label{sec:relative_shift}
The electron-phonon interactions also perturb the energies of the orbital states at second-order. 
The second-order energy shifts $\delta E_-$ (\,$\delta E_+$\,) for states $\ket{e_-}$ (\,$\ket{e_+}$\,) can be expressed in a simple form using the linear phonon $\{x,y\}$ polarisation basis. The energy shift due to phonon modes with wavevector  $k$ and occupation $n_{x(y),k}$ are
\begin{eqnarray}
\delta E_-(x(y),\,k)= \hbar^2\chi_k^2\left(\frac{n_{x(y),\,k}}{\omega-\Delta}-\frac{n_{x(y),\,k}+1}{\omega+\Delta}\right) \nonumber \\
\delta E_+(x(y),\,k) = \hbar^2\chi_k^2\left(\frac{n_{x(y),\,k}}{\omega+\Delta}-\frac{n_{x(y),\,k}+1}{\omega-\Delta}\right)\,,
\end{eqnarray}
where each polarisation contributes independently. Assuming acoustic modes and that the temperatures are such that only modes with frequencies $\Omega\gg\omega\gg\Delta$ contribute significantly to the integral, then correct to lowest order in $\Delta$, the thermal averages of the shifts in the orbital energies over all (acoustic) vibrational levels are
\begin{eqnarray}
\overline{\delta E_-}=\hbar^2\chi\rho\left(-\frac{1}{3}\Omega^3+\frac{\Delta}{2}\Omega^2+\frac{\pi^2k_B^2}{3\hbar^2}T^2\right) \nonumber \\
\overline{\delta E_+}=-\hbar^2\chi\rho\left(\frac{1}{3}\Omega^3+\frac{\Delta}{2}\Omega^2+\frac{\pi^2k_B^2}{3\hbar^2}T^2\right)\,.
\end{eqnarray}
This yields a temperature shift in the spin-orbit splitting
\begin{eqnarray}
\delta\Delta = \overline{\delta E_+}-\overline{\delta E_-}=-\hbar^2\chi\rho\Delta\left(\Omega^2+\frac{2\pi^2k_B^2}{3\hbar^2}T^2\right)\,,
\end{eqnarray}
that is proportional to $T^2$ and a temperature independent mean energy of the orbital states $(\overline{\delta E_+}+\overline{\delta E_-})/2=-\hbar^2\chi\rho\Omega^3/3$.
This correctly predicts the observed $T^2$ dependence of the fine structure splittings in \autoref{fig:optical_transition} (c,d) , but it fails to predict the $T^3$ dependence of the optical line position in \autoref{fig:optical_transition}(b).

\subsection{Optical line position}
\label{sec:mean_shift}
The failure of the above analysis to predict the temperature shift of the optical line position is due to a well known problem in the treatment of the linear Jahn-Teller interaction \cite{englman1972jahn}.
The problem arises from the implicit choice of rectangular mode coordinates for the zero-order vibrational wavefunctions of the perturbative analysis.
In rectangular coordinates, the vibrational wavefunction of a pair ($Q_ x$,$Q_y$) of degenerate $\E$ modes is of the form $\psi_i(Q_x)\psi_j(Q_y)$, where $i$ and $j$ are the independent vibrational quantum numbers of the modes.
Since the rectangular coordinates do not match the cylindrical symmetry of the linear Jahn-Teller vibrational potential, the rectangular vibrational wavefunctions are a poor choice of zero-order basis \cite{longuet1961advances}.
As a consequence, much higher perturbative expansions are required to correctly predict a shift in the optical line position.

A superior choice of basis is obtained by transforming to polar coordinates ($Q_ x$,$Q_y$)$\rightarrow$($\rho$,$\phi$), within which the vibrational wavefunctions take the form $\psi_{\nu,l}(\rho,\phi)$, where $\nu=1,2,\ldots$ is the principal vibrational quantum number and $l=-\nu+1,-\nu+2,\ldots,\nu-1$ is the vibrational angular momentum quantum number, such that the vibrational energies of modes with frequency $\omega$ are $E_{\nu}=\nu\hbar\omega$ \cite{longuet1961advances}.
Using the formalism of the linear Jahn-Teller effect in the polar vibrational basis \cite{longuet1961advances}, we obtained the second-order shifts of the vibrational energies as per Section \ref{sec:relative_shift}.
Performing the thermal average of the shifts in the orbital energies over all (acoustic) vibrational levels, the corrected expression for the temperature shift of the optical line position is
\begin{eqnarray}
\frac{1}{2}(\overline{\delta E_+}+\overline{\delta E_-}) = -2\hbar\chi\rho\int_0^\Omega \frac{2e^{\hbar\omega/k_BT}\left(e^{2\hbar\omega/k_BT}+3\right)}{\left(e^{\hbar\omega/k_BT}-1\right)\left(e^{\hbar\omega/k_BT}+1\right)^2}\omega^2d\omega\propto T^3\,,
\end{eqnarray}
which correctly predicts the $T^3$ dependence of the optical line position.
Note that this corrected approach is consistent with the previous subsection and also predicts a $T^2$ dependence of the fine structure splittings.
Hence, we can conclude that the electron-phonon processes of the linear Jahn-Teller interactions within the ground and excited electronic levels are responsible for the observed temperature variations of the optical line width, position and fine structure splittings.

\section{Discussion}
\label{sec:discussion}

In Section \ref{sec:model}, we have shown that a simple model of linear electron-phonon interactions can be used to successfully explain population dynamics ($\gamma^{e,g}_{+,-}$, Section \ref{sec:first-order_transitions}), dephasing processes ($\gamma^{e,g}_d$, Section \ref{sec:second-order_transitions}), relative (Section \ref{sec:relative_shift}) and mean (Section \ref{sec:mean_shift}) energy shifts within the ground and excited states.
We next discuss implications of our observations for ground state coherences and approaches that could be used to enhance coherence times.

\subsection{Implications for ground state coherences}
\label{subsec:spin_discussion}

\begin{figure}
\hspace{22mm}
\includegraphics{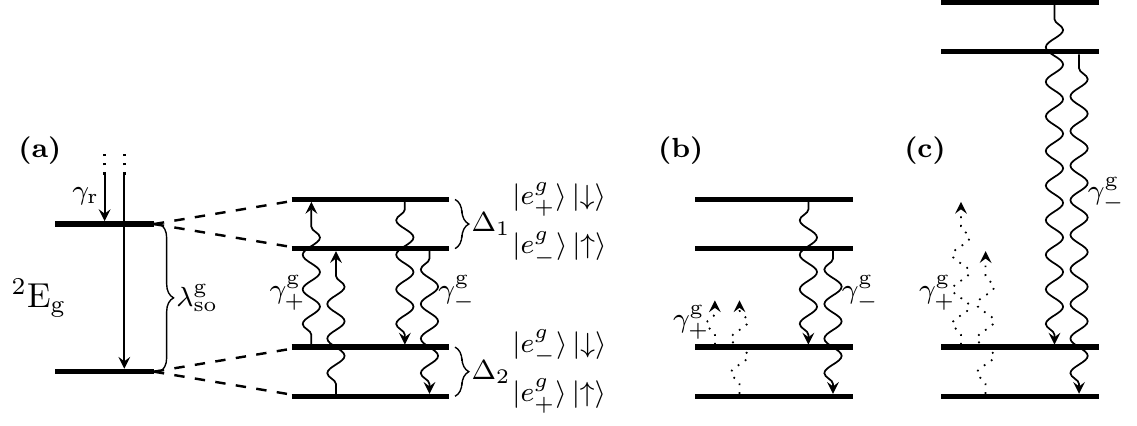}
\caption{
	(a) Implications of phonon processes for ground state coherences.
	The transitions arising from electron-phonon interactions described by our model are spin conserving.
	Any coherences created between two ground states decay with the orbital relaxation rates.
	(b) The $\gamma^{g}_{+}$ can be suppressed at low temperatures ($T \ll \Delta=\lambda^\textrm{g}_{\textrm{SO}}$).
	(c) Large strain fields result in an increased splitting, also resulting in reduced phonon occupation and suppressed $\gamma^{g}_{+}$.
	For (b) and (c), the two lowest energy states constitute a subspace with reduced thermal relaxation and extended coherence.
}
\label{fig:ground_state_spins}
\end{figure}

The $\SiV$ ground states have spin and orbital degrees of freedom which can be used as qubit states.
\autoref{fig:ground_state_spins}(a) shows the electronic states under a magnetic field applied along the $\siv$ symmetry axis.
The orbital relaxation rates $\gamma^{g,e}_{+,-}$ discussed in this manuscript are spin conserving, consistent with the long spin $\Tone$ times that were recently measured \cite{rogers2014all-optical}.
Even though such orbital relaxations are spin conserving, a phonon-induced orbital quantum jump leads to a change of the detuning between spin states.
In this limit, any coherences formed between the four states shown in \autoref{fig:ground_state_spins} are therefore limited by the $\Tone$ of the orbital degree of freedom.

Recent experiments  \cite{pingault2014all,rogers2014all-optical} that probed ground state coherences using coherent population trapping (CPT) have reported $T_2^*$ values that are in good agreement with the orbital $\Tone$ reported in our work.
We note that the model used in \cite{pingault2014all} considered thermal relaxation mechanisms only between the two lowest energy ground states shown in \autoref{fig:ground_state_spins}.
The authors concluded thermal relaxation rates between these two states to be suppressed owing to small spin overlap at low magnetic fields, and the $T_2^*$ to be limited by magnetic field noise from the $\siv$ environment.
While the former agrees with our model (no relaxation between $\ket{e_-}\ket{\downarrow} \leftrightarrow  \ket{e_+}\ket{\uparrow}$), all four ground states need to be considered to relate orbital relaxation rates to coherences.
Based on the close experimental agreement between $T_2^*$ and orbital $\Tone$, we conclude that coherence times of $\siv$ ground states are limited by phonon processes even at liquid helium temperatures.

\subsection{Extending ground state coherences}

We have shown that ground state coherences are limited by a single-phonon orbital relaxation process with a rate determined by a combination of phonon density of states and occupation ($\gamma^g_{\pm}\sim \rho(\Delta)(2\, n(\Delta,T)+1\mp1)$) at the energy of the spin-orbit splitting with $\Delta=\lambdag\sim\SI{50}{\giga\hertz}$.
Since the interaction with the phonon bath is a Markovian process, dynamical decoupling sequences cannot be applied to extend coherences.
To extend $T_2^*$, we will therefore focus on approaches that reduce the orbital relaxation rates $\gamma^g_\pm$.

The first two approaches focus on reducing phonon occupation to decrease $\gamma^g_+$.
The occupation depends on the ratio, $T/\Delta$ , of the temperature and the energy splitting between the coupled orbital states.
Substantial improvements can be achieved by minimizing this ratio in cooling the sample to lower temperatures ($T \ll \Delta \sim \SI{2.4}{\kelvin} $, \autoref{fig:ground_state_spins}(b)).
Based on our fits in Section \ref{sec:exp_ground_states}, the expected orbital relaxation timescale is given by $1/\gamma^g_+=\,101\,(e^{2.4/T}\,-1)\,$\SI{}{\nano\second} which correspond to \SI{1}{\micro\second} at \SI{1}{\kelvin} and \SI{1}{\milli\second} at \SI{0.26}{\kelvin}.
A second approach is to increase $\Delta$ by using emitters subject to high strain.
At the limit of $\Delta \gg T$, similar reductions in phonon occupation can be used to suppress relaxation rates as shown in  \autoref{fig:ground_state_spins}(c).
%
%At 4 K, need 1.6 THz (\[CapitalDelta]\[Lambda] = 2.9 nm) for 1 ms and 0.9 THz (\[CapitalDelta]\[Lambda] = 1.63 nm) strain for 1 us. this corresponds to a wavelength shift of 
Based on Eq. (\ref{eq:firstOrderRate}), we find that $1/\gamma^g_+$ equals \SI{1}{\milli\second} (\SI{1}{\micro\second}) for a strain shift of \SI{1.6}{\THz} (\SI{.9}{\THz}) at \SI{4}{\kelvin}.
We note that in both cases, only the two lowest energy states constitute a subspace that does not couple to phonons.
The lowest two energy states are therefore expected to have long coherence times and could be used as a long-lived spin qubit.

The linear interaction Hamiltonian of Section \ref{sec:electron-phonon_interaction} and the resulting single-phonon orbital relaxation process are analogous to the Jaynes-Cummings Hamiltonian and Wigner-Weisskopf model of spontaneous emission used in quantum optics \cite{yamamoto1999mesoscopic}.
One can therefore use ideas developed in the context of cavity QED to engineer relaxation rates $\gamma^g_\pm$.
In particular, the phonon density of states can be reduced to suppress the orbital relaxation rates.
This is analogous to inhibited spontaneous emission of photons \cite{kleppner1981} which has been observed for microwave and optical photons in atomic and solid-state systems \cite{hulet1985inhibited,lodahl2004controlling,houck2008controlling}.
Acoustic phonons in diamond offer an exciting new platform to probe this effect in a new regime owing to the highly broadband and reflective boundary conditions at the diamond-vacuum interface.
To suppress orbital relaxation rates due to phonons at $\Delta \sim \SI{50}{\giga\hertz}$, small nano diamonds ($d < \SI{120}{\nano\meter}$) can be used to realise a complete phononic band gap for $\nu < \SI{50}{\giga\hertz}$ phonons owing to the strong confinement \cite{albrecht2013coupling}. 
An alternative approach would utilise recent advances in diamond nanofabrication \cite{tao2014single,burek2013nanomechanical} to create 1D-optomechanical structures engineered to inhibit phonon and enhance optical transitions by modifying the density of states \cite{ Safavi2014,Gomis-Bresco2014,kipfstuhl2014modeling,Yablonovitch1987}. 
Using this approach both $\gamma^g_+$ and $\gamma^g_-$ are inhibited, therefore all four ground states can be used as long-lived qubits.
We expect both approaches that modify phonon occupation and density of states to result in substantial improvements for the ground state coherences of $\siv$ centres.

\section*{Acknowledgements}
The authors acknowledge funding from ERC, EU projects (SIQS, DIADEMS, EQUAM), DFG (FOR 1482, FOR 1493 and SFBTR 21), BMBF, USARL/ORISE, DARPA QuASAR, NSF ECCS-1202258, ARC (DP120102232) and Volkswagen foundations for funding.

\section*{Author Contributions}

KJ, AS, LR, JB, and MM performed the experiments, which were conceived by LR, ML, and FJ.
AS, MD, and NM developed the theoretical model.
KJ, AS, and LR wrote the manuscript with input from all the authors.

\section*{References}
\bibliographystyle{unsrt}
\bibliography{electron-phonon_processes}

\end{document}